\documentclass[11pt]{article}
\usepackage{moriond,epsfig}

\def\Journal#1#2#3#4{{#1} {\bf #2}, #3 (#4)}

\def\NIM{\em Nucl. Instrum. Methods}

\def\NPB{{\em Nucl. Phys.} B}
\def\PLB{{\em Phys. Lett.}  B}
\def\PRL{\em Phys. Rev. Lett.}
\def\PRD{{\em Phys. Rev.} D}

\def\be{\begin{equation}}
\def\ee{\end{equation}}
\def\bea{\begin{eqnarray}}
\def\eea{\end{eqnarray}}

\def\tect{$^{130}$Te }

\def\pbdd{$^{210}$Pb }

\def\udt{$^{238}$U }

\def\thdt{$^{232}$Th }

\def\coss{$^{60}$Co }

\def\amnu{$\vert\langle m_{\nu} \rangle\vert$~}

\def\BBz{$\beta\beta(0\nu)$~}

\def\ca{$\sim$}

\def\pom{$\pm$ }

\def\be{\begin{equation}}
\def\ee{\end{equation}}

\def\ciccio{5$\times$5$\times$5 cm$^3$ }
\def\magro{3$\times$3$\times$6 cm$^3$ }

\def\0nbb{DBD0$\nu$}

\begin{document}
\vspace*{4cm}
\title{CUORICINO last results and CUORE R\&D}
\author{ 
S.~CAPELLI$^{1}$,
R.~ARDITO$^{1,2}$,
C.~ARNABOLDI$^{1}$,
D.~R.~ARTUSA$^{3}$,
F.~T.~AVIGNONE~III$^{3}$,
M.~BALATA$^{4}$,
I.~BANDAC$^{3}$,
M.~BARUCCI$^{5}$,
J.W.~BEEMAN$^{6}$,
F.~BELLINI$^{14}$,
C.~BROFFERIO$^{1}$,
C.~BUCCI$^{4}$,
F.~CAPOZZI$^{1}$,
L.~CARBONE$^{1}$,
S.~CEBRIAN$^{7}$,
M.~CLEMENZA$^{1}$,
C.~COSMELLI$^{14}$,
O.~CREMONESI$^{1}$,
R.~J.~CRESWICK$^{3}$,
I.~DAFINEI$^{14}$,
A.~DE~WAARD$^{8}$,
M.~DIEMOZ$^{14}$,
M.~DOLINSKY$^{6,11}$,
H.~A.~FARACH$^{3}$,
F.~FERRONI$^{14}$,
E.~FIORINI$^{1}$,
G.~FROSSATI$^{8}$,
C.~GARGIULO$^{14}$,
E.~GUARDINCERRI$^{10}$,
A.~GIULIANI$^{9}$,
P.~GORLA$^{7}$,
T.D. GUTIERREZ$^{6}$,
E.~E.~HALLER$^{6,11}$,
I.~G.~IRASTORZA$^{7}$,
E.~LONGO$^{14}$,
G.~MAIER$^{2}$,
R. MARUJAMA$^{6,11}$,
S.~MORGANTI$^{14}$,
S.~NISI$^{4}$,
C.~NONES$^{1}$,
E.~B.~NORMAN$^{13}$,
A.~NUCCIOTTI$^{1}$,
E.~OLIVIERI$^{5}$,
P.~OTTONELLO$^{10}$,
M.~PALLAVICINI$^{10}$,
V.~PALMIERI$^{12}$,
E.~PASCA$^{5}$,
M.~PAVAN$^{1}$,
M.~PEDRETTI$^{9}$,
G.~PESSINA$^{1}$,
S.~PIRRO$^{1}$,
E.~PREVITALI$^{1}$,
B. QUINTER$^{6,11}$,
L.~RISEGARI$^{5}$,
C.~ROSENFELD$^{3}$,
S.~SANGIORGIO$^{9}$,
M.~SISTI$^{1}$,
A.~R.~SMITH$^{6}$,
S.~TOFFANIN$^{12}$,
L.~TORRES$^{1}$,
G.~VENTURA$^{5},$
N.~XU$^{6},$
~and~L.~ZANOTTI$^{1}$
 }

\address
 {1. Dip. di Fisica dell'Univ. di 
Milano-Bicocca e Sez. di Milano dell'INFN, Milano I-20126,
Italy\\
 2. Dip. di Ingegneria Strutturale del Politecnico di
Milano, Milano I-20133, Italy\\
 3. Dept.of Physics and Astr.,
University of South Carolina, Columbia, South Carolina, USA 29208\\
4. Laboratori Nazionali del Gran Sasso, I-67010,
Assergi (L'Aquila), Italy\\
 5. Dip. di Fisica dell'Univ. di
Firenze e Sez. di Firenze dell'INFN, Firenze I-50125, Italy\\
 6. Lawrence Berkeley National Laboratory, Berkeley,
California, 94720, USA\\
 7. Laboratorio de Fisica Nuclear y Altas Energias,
Universid\`{a}d de Zaragoza, 50009 Zaragoza, Spain\\
 8. Kamerling Onnes Laboratory, Leiden University,
2300 RAQ, Leiden, The Netherlands\\
 9. Dip. di Fisica e
Matematica dell'Univ. dell'Insubria e Sez. di Milano
dell'INFN, Como I-22100, Italy\\
10. Dip. di Fisica dell'Univ. di
Genova e Sez. di Genova dell'INFN, Genova I-16146, Italy\\
11. University of California, Berkeley, California
94720, USA\\
12. Laboratori Nazionali di Legnaro, 
I-35020 Legnaro ( Padova ), Italy\\
13. Lawrence Livermore National Laboratory, 
Livermore, California, 94550, USA\\
14. Dip. di Fisica dell'Univ. di Roma e Sez. di Roma 1 dell'INFN, Roma  I-16146, Italy
}

\maketitle\abstracts{
	CUORICINO is a bolometric experiment on Neutrinoless Double Beta Decay (DBD0$\nu$) 
of \tect. It consists of an array of 62 TeO$_ 2$ crystals with a total mass of $\sim$ 40.7 kg.
While being a self consistent experiment CUORICINO is also a good test 
for the feasibility of the next generation experiment CUORE, \ca 750 kg of TeO$_ 2$ bolometric mass.
	In this paper last results from CUORICINO and prospects for the future CUORE 
experiment will be reported.}

\section{Introduction}
The positive results obtained in the last few years in neutrino oscillation experiments ~\cite{SK+,SNO+,SNO04,kaml03,K2K03}
have given convincing and model indipendent evidences that neutrinos are massive and mixed particles. The obtained data 
are compatible with two possible mass patterns, or hierarchies, the normal: $m_1 < m_2 << m_3$, and the inverted hierarchy: 
$m_3 << m_1 < m_2$. 
In this scenario there are two general theoretical possibilities for the neutrino with definite mass, depending on the
conservation or not of the total lepton charge. In the first case neutrinos are Dirac particles, while in the 
second case they are Majorana particles.
Unforunately oscillation experiments are only sensitive to neutrino mass eigenvalue differences squared, but cannot give any 
information with respect to neutrino nature and absolute mass scale ~\cite{bilenky80,langacker87}. 

Experiments looking for the DBD$0\nu$ of even-even nuclei have the highest sensitivity to possible violations 
of the total lepton number L and to Majorana neutrino masses. A positive signal would therefore give a clear answer with respect
to neutrino nature and absolute mass scale. 

In this lepton violating  process, a nucleus (A,Z) decays into (A,Z+2) with the emission of two electrons and no neutrino. 
This leads to a peak in the sum energy spectrum of the two electrons at the Q-value of the transition. 
The decay rate of this process is given by the equation:

\begin{equation}\label{eq:t0n} %
\left[ T _{1/2}^{0\nu } \right]^{-1} = \mid M^{0\nu}\mid^2~G^{0\nu}~\frac{\mid <m_\nu>\mid^2}{m_e^2}  
\end{equation}

\noindent where  $G^{0\nu}$ is the two body phase-space factor including coupling costants and $ M^{0\nu}$ is 
the nuclear matrix element.

The quantity  $\mid <m_\nu>\mid$ is the effective electron neutrino  Majorana mass 
which can be expressed in terms of the elements of the neutrino mixing matrix as follows:
\begin{equation}\label{eq:meff}
\vert\langle m_{\nu} \rangle\vert \equiv \vert \vert
U_{e1}^L \vert ^2m_1 + \vert U_{e2}^L \vert ^2m_2 e^{i\phi _2 } +
\vert U_{e3}^L \vert ^2m_3 e^{i\phi _3 }\vert ,
\end{equation}

\noindent where $e^{i\phi _2 }$ and $e^{i\phi _3 }$ are the Majorana CP--phases (\pom1 for CP conservation), 
$m_{1,2,3} $ are the Majorana neutrino mass eigenvalues and  U$^L_{ej}$ are the coefficients of the 
Pontecorvo-Maki-Nakagawa-Sakata (PMNS) neutrino mixing matrix, determined from neutrino oscillation data. 

Recent global analyses of all oscillation experiments ~\cite{Feruglio,Ferugliobis,Pascoli,Pascolibis} 
yield on average:

\begin{eqnarray}\label{eq:meffexp} 
\vert\langle m_{\nu} \rangle\vert = \vert (0.70\pm0.03) m_1 + (0.30\pm 0.03) m_2 e^{i\phi _2 } + (<0.05) m_3 e^{i\phi _3 }\vert 
\end{eqnarray}

As it can be seen from equation (\ref{eq:t0n}) any uncertainty on the nuclear matrix element $M^{0\nu}$ reflects 
directly on the measurement of $\mid <m_\nu>\mid$. 
Many calculations exist in literature \cite{faessler98,suhonen98,elliot02,civitarese03} and find 
rather different nuclear matrix element values. This is one of the regions for which it 
is fundamental to search for DBD0$\nu$ on several nuclei \cite{bilenky04}.

No evidence for DBD0$\nu$ has been reported so far, with the exception of the claimed  
discovery of the decay of $^{76}$Ge from which a best fit value for \amnu of 0.4 eV (99.9973\% C.L.) 
is given \cite{Klapdor,klapdor04,Klapdorter}.
Actually the best lower bound on  $\mid <m_\nu>\mid$ is the one obtained by the Heidelberg-Moscow experiment
 with $Ge^{76}$ and is $\mid <m_\nu>\mid ~\leq$~ (0.1-0.9) eV (99.73\% C.L.) \cite{klapdor04}.

 DBD$0\nu$ can be searched for with different experimental methods. One possible direct approach is based on the
 bolometric technique \cite{twerenbold96}. The energy released in dielectric and diamagnetic crystals gives rise to measurable 
 temperature increases when working at low temperature (T \ca 10 mK). The absorber material 
 can be chosen quite freely, the only requirements being, in fact, reasonable thermal 
and mechanical properties. The absorber can therefore be easily built with materials containing 
any kind of unstable isotopes as for instance DBD candidates.
 This method has the same advantages of solid state detectors (high detector mass and good energy resolution) 
 and offers at the same time a wide choice of DBD candidates to be used.
 The isotope \tect is and excellent candidate to search for DBD due to its high transition energy 
 (2528.8 $\pm$1.3 keV) and large isotopic abundance (33.8\%) which allows a sensitive experiment 
 to be performed with natural tellurium. Of the various compounds of this element, TeO$_2$ appears to be the most
 promising, due to its good thermal and mechanical properties.
 
 Because of the rarity of the searched process, spurious counts due to environmental and cosmic 
 radioactivity, airbone activity (Rn), neutrons, intrinsic contaminations and cosmic ray activation 
 of the detector and of the other experimental setup materials, can obscure the signal counts of interest.
 A good knowledge of the radioactive sources that mainly 
 contribute to the measured background in the DBD$0\nu$ energy region, is therefore of fundamental importance
 in order to study new strategies to reduce such contaminations and consequently improve the sensitivity of 
 the experiment.
  
The first TeO$_2$ large mass bolometer (MiDBD experiment) consisted in an array of 20 340 g natural TeO$_2$ crystals 
for a total mass of 6.8 kg. 
It was operated between 1997 and 2001 in the hall A of the National Laboratories of Gran Sasso (LNGS), that consitute 
a fundamental shield against Cosmic Rays (3200 m.w.e.).
 MiDBD collected a total statistics of 4.3 kg$\cdot$y and measured a background in the 
 DBD0$\nu$ region of 0.3 c/keV/kg/y. The reached limit for the halflife of 
 \tect as respect to DBD0$\nu$ is $2.08\cdot 10^{23}$y (90$\%$ CL) \cite{arnaboldi03}.\\
 After some tests with larger crystals ($\sim$ 750 g each) in 2001 this experiment was dismounted in order
 to leave the place to the larger mass TeO$_2$ experiment CUORICINO.  

 \begin{figure}[t!bp]
\begin{center}
 \includegraphics[height=9cm]{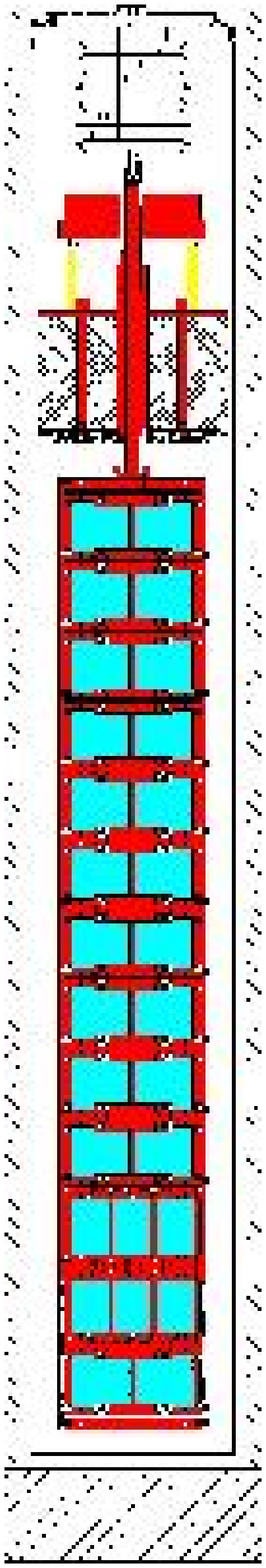}
 \hskip 2.5 cm 
 \includegraphics[height=9cm]{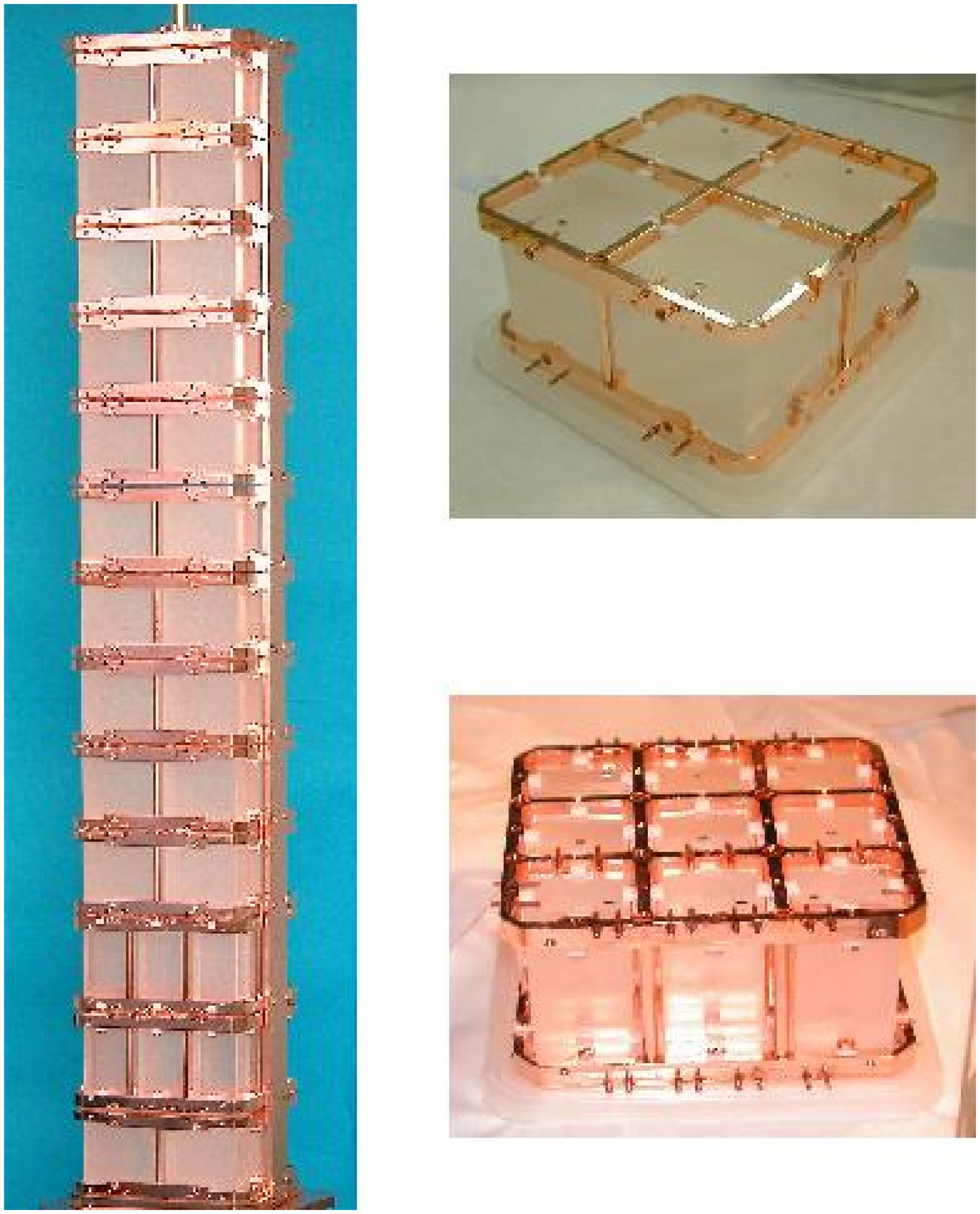}
 \end{center}
 \caption{\small{\em{The CUORICINO detector: scheme of the tower and internal roman lead shields (left), 
 the 13 planes tower (centre), the 4 crystal module (top right) and the 9 crystal module (bottom right).}}}
 \label{fig:cuoricino}
\end{figure}

\section{CUORICINO set-up}
CUORICINO is a tower-like structure made by eleven modules of 4 detector each (5x5x5 cm$^3$ 790g TeO$_2$ crystals)
and two modules of 9 detector each (3x3x6 cm$^3$ 330g TeO$_2$ crystals) for a total mass of about 
40.7 kg (see fig.~\ref{fig:cuoricino}).
The 18 small size crystals were taken from the previous MiDBD experiment.
All the crystals are made of natural tellurium but the 4 isotopically enriched crystals previously 
used in MiDBD.

The experience gained with MiDBD made us aware of the necessity of working with materials with an intrinsic 
and surface contaminations reduced as much as possible. Particular care was therefore devoted to selection 
and cleaning of the CUORICINO detector materials. The 5x5x5 cm$^3$ TeO$_2$ crystals were grown 
from pre-tested low activity powders and all the crystals were subject to surface treatements with low activity
materials. The mechanical structure of the array was made exclusively in OFHC copper and PTFE, both 
previously measured in order to make sure of their extremely low radioactive content. 
All the copper and PTFE parts facing the detectors were 
separately treated with acids in order to remove any possible surface contamination.
All the detector mounting operations were performed  in an underground clean room in a
 N$_2$ atmosphere to avoid Rn contaminatin. 
The tower just after assembly completion and the details of the two different 
used modules are shown in fig.~\ref{fig:cuoricino}.
 
The detector tower is surrounded by copper plates and installed in the same $^3$He-$^4$He 
dilution refrigerator previously used for MiDBD, at a working temperature of about 10 mK. 
Roman lead shields are placed all around the detector in order to avoid radioactivity
coming from the cryostat. Due to the bigger dimension of the CUORICINO tower with respect to MiDBD,
 the amount of Roman lead surrounding the CUORICINO tower
is less than before: $\sim$ 1 cm thick roman lead shield (vs. 3 cm in MiDBD) 
is placed on the side of the detector tower and two roman lead discs of 7.5 cm (vs. 10 of MiDBD) 
and 10 cm thickness (vs. 15 of MiDBD) respectively, 
are positioned just below and above the tower respectively.
The tower is mechanically decoupled from the cryostat through a steel 
spring in order to avoid vibrations from the overall facility to reach the detectors. 

The dilution refrigerator is shielded against environmental radioactivity by two layers of 
 lead of 10 cm minimum thickness each. The outer layer is of commercial low radioactivity lead,
  while the internal one is made with special lead with a \pbdd contamination of 16 $\pm$ 4 Bq/kg. 
The external lead shields are surrounded by an air-tight box flushed with fresh nitrogen to avoid radon 
contamination to reach the detector. 
A borated polyethylene neutron shield (10 cm) is also present.
All the structure is housed inside Faraday Cage in order to suppress electromagnetic interferences. 

\section{CUORICINO performances and results}
CUORICINO first measurement started in March 2003. Unfortunately some detector connections broke during the
cooling down procedure, so that only 32 \ciccio and 17 \magro crystals could be read. 
Since the active mass was anyway quite large (\ca 30 kg of TeO$_2$) and the detector performances
 were quite good data collection was continued for a few months.
The average pulse height obtained with the working detectors is of 120 $\pm$ 75 $\mu$V/MeV*kg for the \ciccio
crystals and 104 $\pm$ 35 $\mu$V/MeV*kg for the \magro crystals. The average 
resolution FWHM in the DBD0$\nu$ region was evaluated on the 2615 keV $^{208}$Tl line 
measured during calibration with a \thdt source. It is 7.8 $\pm$ 2.8 keV for the bigger size and of 
9.1 $\pm$ 3.1 keV for the small size crystals.

At the end of October 2003 CUORICINO was stopped to undergo substantial operations of maintenance and to 
recover the lost electrical connections and hence increase the number of working detectors.
At the end of April the second run of CUORICINO started. Unfortunately two of the \ciccio 
crystal wire connections had broken during the cooling down procedure. Actually we have 42 big crystals over 44 and 
all the 18 small crystals working with a TeO$_2$ active mass of about 39 kg.
The average pulse height obtained with the working detectors is of 167 $\pm$ 99 $\mu$V/MeV*kg for the \ciccio
crystals and 147 $\pm$ 60 $\mu$V/MeV*kg for the \magro crystals. The average 
resolution FWHM is 7.5 $\pm$ 2.9 keV for the bigger size and of 
9.6 $\pm$ 3.5 keV for the small size crystals.

\begin{figure}[t!]
\begin{center}
 \includegraphics[height=8cm]{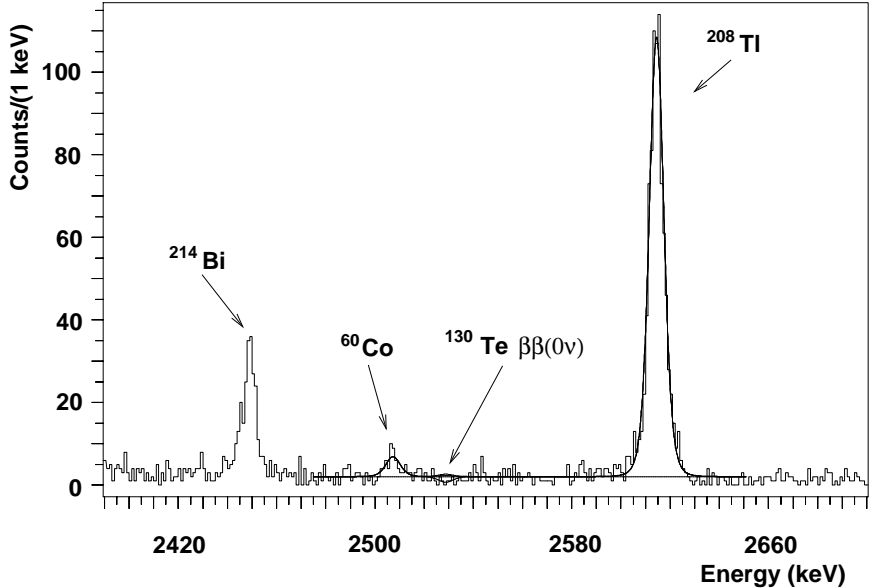}
\caption{\label{fig:Spectrum} Spectrum of the sum of the two electron energies in
the region of neutrinoless DBD}
\end{center}
\end{figure}

The results presented here refer to the data acquired with the working TeO$_2$ crystals 
up to December 2004, with a total statistics of 10.85 kg*y.
The sum of the spectra of the \ciccio and \magro crystals in 
the DBD0$\nu$ region is shown in fig.~\ref{fig:Spectrum}, where the peaks at  
2447, 2505 keV and 2615 keV due to \udt, \coss and \thdt contaminations respectively are clearly visible.

The background value measured in this region is of 0.18 $\pm$ 0.01 c/keV/kg/y, 
a factor of about 2 less than the one obtained in the previous experiment MiDBD, thus indicating the effectiveness of
the cleaning operation performed in CUORICINO. 
 No evidence is found for a peak at 2528.8 keV, the energy of the DBD0$\nu$ of the isotope \tect.
By applying a maximum likelihood procedure \cite{Baker,Barnett} to search for the maximum signal 
compatible with the measured background, we obtain a 90\% C.L. lower limit of 1.8 $\times$ 10$^{24}$ y 
on the \tect lifetime for this decay. 
This limit leads to a
constraint on the electron neutrino effective mass ranging from 0.2 to 1.1 eV, depending on the nuclear 
matrix elements considered in the computation \cite{arnaboldi05}. By using the same nuclear matrix element value used 
by  H.V. Klapdor-Kleingrothaus et al.~\cite{Klapdorter} we obtain a value for \amnu of 0.527 eV.
The reported data show that CUORICINO is a competitive experiment in the field of Neutrinoless 
Double Beta Decay \cite{arnaboldi04}. It has a 5 year sensitivity 
(at 68\% C.L.) of about 9 x 10$^{24}$ year for the  \BBz  of \tect. This means that CUORICINO will be able 
to test the Majorana mass in the 100-700 meV range.

\section{From CUORICINO to CUORE}

The good results and performances obtained with CUORICINO look very promising in view of the future experiment
CUORE. It has in fact demonstrated the feasibility of a large bolometric array of TeO$_2$ crystals in a
tower-like structure and at the
same time it has shown that detector performances are not affected by the increase in crystal size (from 330 g to
790 g).

CUORE will be a tightly closed structure of 988 TeO$_2$ \ciccio crystals arranged in 19 CUORICINO-like 
towers, for a total mass of \ca 741 kg.
It will be provided with copper shields, internal roman lead shields (2 cm thick all around the detector 
plus 20 cm thick on the top), external commercial lead shields (20 cm thick) and neutron shield (10 cm thick).
The cryostat will be kept in nitrogen overpressure to avoid radon contamination from the air to reach the detector
and placed in a Faraday Cage in order to suppress electromagnetic intereferences.
The goal of this experiment is to reach a background rate in the DBD0$\nu$ energy region in the range 0.001 to
0.01 c/keV/kg/y, corresponding to a sensitivity on the effective electron neutrino mass 
of 0.02--0.07~t$^{-1/4}$ eV and 0.03--0.13~t$^{-1/4}$ eV respectively.

Unfortunately CUORICINO can't be a direct test for what concerns CUORE background, due to the different detector
geometry and materials that will be used.
Indeed the tightly closed structure of the CUORE detector will allow a strong background suppression working with
all the detectors in anticoincidence.
Moreover the lead shield designed for CUORE will be optimized in order to practically cancel the background 
coming from outside. This optimization was not possible in CUORICINO housed in the old cryostat used for 
MiDBD where the space is limited.
The intense R\&D activities as respect to surface cleaning and material selection will also give an additional
reduction in the background contribution to the DBD0$\nu$ region.

The background results obtained with CUORICINO are very promising. They demonstrate that our knowledge
of the main background sources and the efforts made to reduce them are in the good direction. 

\begin{figure}[t!]
\begin{center}
 \includegraphics[height=9cm]{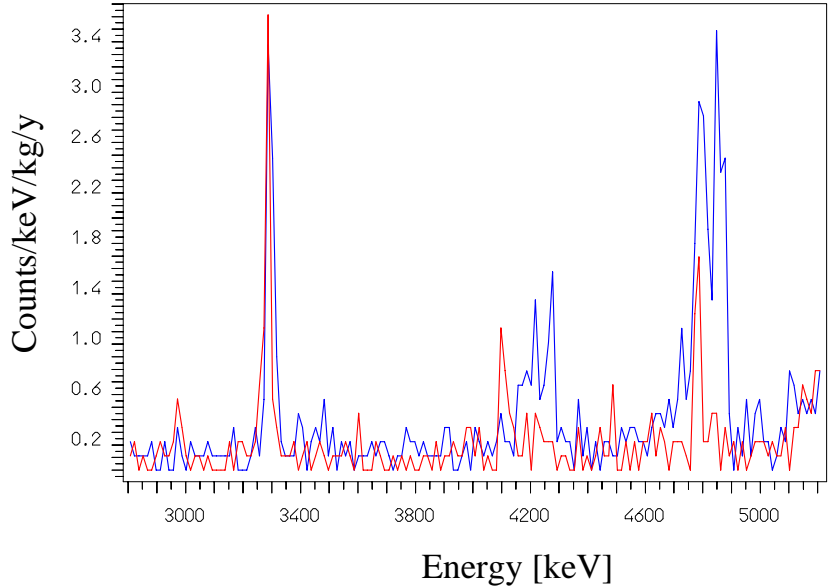}
\caption{\label{fig:salaC} Comparison betweeen the background spectra obtianed in Cuoricino (top
and bottom planes) (blu) and with the test set-up (red).}
\end{center}
\end{figure}

A dedicated study of the CUORICINO background was performed both with sophisticated analysis procedures and with
the aid of Montecarlo simulations of the whole detector set-up. 
A background model, able to describe the observed measured spectra  
in terms of environmental radioactivity, radioactive bulk contaminations of the whole detector setup 
and surface contaminations of the materials directly facing the detector itself, was developed \cite{capelli-nu04}.
The sources identified as possible responsible for the measured background in the DBD0$\nu$ region are $\beta$ and
$\alpha$ decays from the surface of the crystals and of the parts directly facing them (the biggest is due to 
copper) and multi-Compton events from $^{208}$Tl gamma decays, probably due to \thdt contaminations of 
the materials far away from the detectors.
The evaluated surface contamination level for both the crystals and the copper facing them is around 10$^{-9}$
g/g, leading to contributions to the DBD0$\nu$ energy region of the order of 20~$\pm$~10\% and 50~$\pm$~20\%
respectively. 
The evaluated contribution due to \thdt contaminations of distant parts is of $\sim$ 30~$\pm$~10\%.

The results obtained with this analysis have demonstrated to be very useful both in addressing the R\&D effort 
with respect to material selection and surface cleaning and in allowing an evaluation of the background reachable
in CUORE with the contamination levels measured so far for the materials presently at our disposal.
With the aid of Montecarlo simulations of the entire CUORE geometry and using the same background model tested on
CUORICINO, the contribution expected from bulk and surface contaminations of the crystals and of the copper at the level
measured in CUORICINO have been evaluated.
For \thdt and \udt bulk contaminations at levels of about 10$^{-13}$g/g and 10$^{-12}$g/g for the crystals and the copper 
respectively, the evaluated contribution to the DBD0$\nu$ energy region in CUORE is less than 2$\times$10$^{-2}$ c/keV/kg/y.
The evaluated contribution due to surface contamination levels  of about 10$^{-9}$g/g for both crystals and copper 
is around 7$\times$10$^{-2}$ c/keV/kg/y.
Reductions of factors at least 4 and 10 for surface contamination of crystals and copper respectively
 are necessary in order to reach the wanted sensitivity with CUORE. 

An intense R\%D activity with respect to surface cleaning optimization and measurement is under progress.
New cleaning procedures for both crystals and copper surfaces have been tested in January 2005 
in a second cryostat installed in the hall C of LNGS, provided with copper and lead shields analogous to that used in
 CUORICINO \cite{pirro05}.
The detector consisted in 8 \ciccio TeO$_2$ crystals, arranged in 2 planes of 4 detectors each.  
The crystal surfaces were treated with a nitric acid solution and successively polished with selected high-purity SiO$_2$ lapping 
powders. The copper structure was subject to electro-erosion in an ultra-clean solution of citric acid.
All the assembling precedure was performed in clean environment and using  materials with measured low radioactive content.

The energy spectrum obtained with the collected data shows a reduction with respect to CUORICINO (fig. \ref{fig:salaC}) 
in the counting rate in the region above 4 MeV, where the main contribution is due to alpha decays occurring in the crystals bulk and surface and on the surface of material
facing them (most likely copper).
In particular a reduction of a factor of about 4 is observed in the peaks at the transition energy of \udt and \thdt alpha decays.
According to the background model previously described and updated to the actual detector geometry, 
these peaks can be attributed to contaminations in the first $\mu$m depth of the crystal surface. 
The result obtained in this test funfilles therefore the CUORE requirement with respect to crystal surface contamination level.
As regards copper no indication of surface contamination reduction has been observed.
  
In order to exclude possible contributions to the DBD0$\nu$ energy region arising from small components facing the crystals (i.e. PTFE 
PTFE parts, silicon heaters and gold bonding wires) a new test has been performed with the same detector used in the
 previous test measurement.
The crystal surfaces have been covered with large amounts of the components to be tested in order to have a high sensitivity. Preliminary results 
give indications of very small contributions due to these elements, leaving as the most probable background source
 the copper surface.

New and more effective cleaning procedures for the copper surface are therefore mandatory. 
Recent measurement performed with laser ablation have given positive indications in favour of the plasma cleaning technique
\cite{palmieri05}.

\section{Conclusions}

CUORICINO ($\sim$ 41 kg of TeO$_2$ crystals) is a sensitive experiment on Neutrinoless Beta Decay able to reach in 5 years a limit on the effective electron neutrino mass
ranging from 0.2 to 1.1 eV, depending on the nuclear matrix elements considered in the computation.
While being a self consistent experiment CUORICINO is also a fundamental test for the next generation CUORE experiment 
($\sim$740 of TeO$_2$ crystals) for what concerns detector performances and background value in the DBD0$\nu$ region.

The good results obtained with CUORICINO have shown the feasibility of the tower-like structure and have demonstrated that
detector performances are not affected by the mass increasing from 330 g to 790 g.
The results obtained with CUORICINO allowed us also to make an evaluation of the background attainable in CUORE with the
contamination levels measured for the materials actually at our disposal. 
Reduction factors of about 4 and 10 of crystals and copper surface contaminations are necessary in order 
to reach the wanted sensitivity. An intense R\%D with respect to material selection and surface cleaning is under progress.
Recent test and measurements have given promising results.

For a background value in the DBD0$\nu$ region of about 0.001 c/keV/kg/y CUORE could reach a sensitivity for \amnu
in the range 30--100 meV in 5 years, just in the region favoured by current oscillation experiments for the inverted hierarchy.

\end{document}